# HyGAS: an Open, Sensor-Agnostic Platform for Multi-Satellite Methane Plume Retrieval, Uncertainty Propagation, and Emission-Rate Estimation

Alvise Ferrari
*GMATICS S.r.l.*
Rome, Italy
*School of Aerospace Engineering*
*Sapienza University*
Rome, Italy
alvise.ferrari@uniroma1.it

Valerio Pampanoni
*School of Aerospace Engineering*
*Sapienza University*
Rome, Italy
valerio.pampanoni@uniroma1.it

Giovanni Laneve
*School of Aerospace Engineering*
*Sapienza University*
Rome, Italy
giovanni.laneve@uniroma1.it

***Abstract*— The rapid expansion of spaceborne methane observing capabilities at the facility-scale (fostered both by public missions and commercial constellations) has created a need for harmonised, reproducible, and uncertainty-aware processing chains that support both monitoring workflows and fair inter-sensor comparisons. This paper presents HyGAS (Hyperspectral Gas Analysis Suite), an open and sensor-agnostic framework that standardises methane processing across multiple imaging spectrometers. HyGAS currently supports end-to-end processing from Level-1 radiance to methane enhancement for PRISMA, EnMAP, and Tanager-1, and it supports ingestion of Level-2 methane enhancement products from EMIT and GHGSat, which are subsequently processed through common downstream modules for background selection, plume segmentation, Integrated Mass Enhancement (IME), and emission-rate inversion. HyGAS prioritises operational robustness via (i) matched-filter variants designed to mitigate background heterogeneity and pushbroom artefacts, (ii) explicit decomposition and propagation of uncertainty from instrument noise and scene-driven clutter to IME and flux, and (iii) a scale-aware segmentation strategy defined in physical units and rescaled by ground sampling distance to improve multi-sensor comparability. Representative sample outputs are reported for PRISMA, EnMAP, and Tanager-1.**

## I. Introduction

Methane is a powerful greenhouse gas, and a substantial fraction of emissions is associated with spatially localised "super-emitters" (e.g., landfills and fossil-fuel infrastructure) that can be detected and quantified with imaging spectroscopy [1], [2], [3]. The last few years have seen a fast growth of spaceborne and commercial imaging spectrometers capable of facility-scale methane mapping, including PRISMA [4], EnMAP [5], EMIT [6], and GHGSat [7], with emerging contributions from new missions such as Tanager-1 [8], [9]. This growth is enabling broader monitoring use cases, including global identification of key source categories (e.g., solid waste) using hyperspectral satellites [10], and community efforts to assess performance across multiple satellite systems [11], [12].

Despite this progress, exploiting the multi-sensor ecosystem as a coherent observing system remains difficult. Sensors differ in spectral response functions (SRF), spatial resolution, noise characteristics, and calibration artefacts typical of pushbroom imaging spectrometers [13], [14]. Meanwhile, methane retrieval approaches span Matched-Filter (MF) families and other paradigms, and even when different sensors observe the same source, differences in background handling, segmentation choices, and uncertainty treatment can dominate the spread in IME and flux estimates [3], [15], [16].

Uncertainty is a central blocker for operational decision-making and fair intercomparison. Methane enhancement precision is controlled by instrument noise (often summarised by SNR and dominated by photon statistics at high-enough levels of radiance [17]), surface-driven spectral variability ("clutter"), and structured artefacts such as striping and smile that can leak into retrieval outputs if not accounted for in the background model [13], [18]. While analytical uncertainty propagation exists for MF retrievals and is widely used in plume-to-flux quantification [3], consistent treatment across sensors is still not common practice.

This paper presents HyGAS (Hyperspectral Gas Analysis Suite) as a practical response: a unified, open framework that enables (i) sensor-agnostic methane enhancement retrieval where Level-1 radiances are available, (ii) standardised downstream processing (background selection, segmentation, IME, flux), and (iii) explicit uncertainty propagation from enhancement to flux to support monitoring and multi-sensor comparative studies. The HyGAS codebase is publicly available, the software is free and open-source, and distributed under a GPL-3.0 license [19].

Platform contributions emphasised here:

- A modular chain supporting end-to-end workflows for PRISMA/EnMAP/Tanager-1 and harmonised ingestion for EMIT/GHGSat [4]–[6], [8], [7], [9], [20].
- Operationally robust MF variants (scene-wide, clustered, column-wise) designed for heterogeneous scenes and pushbroom artefacts while preserving physically interpretable enhancement units [21], [15], [16].

- A first-class uncertainty model separating instrument noise and scene-level clutter, propagated through enhancement → IME → flux [3], [17], [18].
- A scale-aware segmentation approach intended to reduce subjectivity and improve cross-sensor comparability (described here from an operational standpoint and fully detailed in the extended HyGAS paper [22]).

## II. HyGAS SCOPE AND DESIGN PRINCIPLES

HyGAS is designed as a sensor-agnostic methane analysis platform**:** the same conceptual pipeline is applied across missions, while sensor-specific elements are encapsulated in instrument adapters (e.g., SRF/smile handling, radiometric noise characterisation, and product mapping). At the time of writing, HyGAS supports end-to-end processing from Level-1 radiance to MF methane enhancement for PRISMA, EnMAP, and Tanager-1 [4], [5], [8], [9]. It also ingests Level-2 enhancement products (with associated uncertainty fields where available) for EMIT and GHGSat [6], [7], [23], [20], enabling a common downstream chain for background selection, segmentation, IME, and emission-rate estimation [3]. Additionally, support for EMIT Level-1 radiance products is expected to be implemented in the short term.

A core design choice is that HyGAS outputs are physically interpretable and decision-ready: MF retrievals are implemented to output methane enhancement maps in concentration–path-length units (ppm·m), enabling consistent IME and flux estimation without ad-hoc empirical scaling [3], [21], [15]. This choice is aligned with the broader MF methane literature and with existing spaceborne methane mapping demonstrations over PRISMA and EnMAP [24], [25].

HyGAS also formalises the idea that multi-sensor monitoring is not just a retrieval problem: it requires coherent decisions about how "background" is defined and sampled, how plumes are segmented across different pixel sizes, and how uncertainty is propagated so that differences between sensors can be interpreted quantitatively [3], [15], [16].

## III. CORE CAPABILITIES (OPERATIONAL VIEW)

### A. *End-to-end methane retrieval with matched-filter variants*

For sensors with accessible Level-1 radiances (PRISMA, EnMAP, Tanager-1), HyGAS implements a MF framework for methane enhancement retrieval, drawing on established imaging-spectroscopy methane mapping approaches including cluster-tuned matched filtering [21] and related retrieval practice in airborne/spaceborne imaging spectroscopy [27].

The central operational requirement is robustness in the presence of (i) heterogeneous land cover and illumination, (ii) pushbroom artefacts that vary across track [13], [18], and (iii) SRF variability and spectral distortions that can bias methane targets and background models if not accounted for [15]. HyGAS therefore provides multiple retrieval configurations that share a common statistical foundation but differ in how background statistics are estimated:

- Scene-wide CMF, for stable global background statistics in homogeneous conditions;
- CTMF (cluster-tuned MF), to improve performance in scenes with multiple surface types by estimating cluster-specific background statistics [21];
- CWCMF (Column-wise CMF), to explicitly adapt background mean/covariance per detector column, improving stability for pushbroom artefacts and across-track variability (particularly important for monitoring and threshold-based plume delineation).

HyGAS also supports multi-configuration retrieval runs to quantify sensitivity to retrieval settings and scene-specific enhancement spectra, which are known to influence MF response and detection/quantification performance [15]. Finally, HyGAS follows the well-known MF caution regarding signal contamination in background statistics; this motivates robust background selection and, where appropriate, localised estimation strategies [16].

### B. *Uncertainty as a first-class product (from enhancement to flux)*

HyGAS treats uncertainty as a primary output required for monitoring and intercomparison. For Level-1-based retrievals, HyGAS propagates instrument noise through the MF to obtain a per-pixel enhancement precision layer. Instrument noise is driven by photon statistics at sufficient radiance and by additive terms (e.g., dark/read components), consistent with standard radiometric noise models [17]. In practical mission contexts, noise characterisation can be supported by mission- and sensor-specific calibration/performance studies (e.g., EMIT on-orbit calibration and performance) [14] and by mission-specific noise coefficient retrieval approaches (e.g., for PRISMA) [18].

HyGAS explicitly decomposes enhancement variability into two interpretable components:

- Instrument noise, propagated analytically through the MF (Level-1 retrievals);
- Scene-driven clutter, representing residual enhancement variability due to surface spectral variability and imperfect modelling, estimated from spectrally matched plume-free pixels (Section 3.3).

These components are combined into a total $\sigma(\Delta X)$ and aggregated within plume masks to obtain IME uncertainty. Emission rates are then derived using an IME-based inversion and uncertainty propagation that combines IME and wind terms [3]. Wind speed can be taken from reanalysis products such as ERA5 in many operational settings [25], with the chosen uncertainty treatment reported alongside flux results [3].

For EMIT and GHGSat, HyGAS can ingest enhancement and uncertainty fields from official/proprietary processors and carry them through the same plume/IME/flux modules, enabling harmonised downstream analysis even when Level-1 retrieval is not performed within HyGAS [7], [23], [20].

### C. *Spectrally matched background selection*

A recurring operational pain point in methane mapping is background definition: manual "background boxes" near the plume can strongly bias covariance estimation and clutter quantification, particularly in heterogeneous scenes and for MF retrievals [15], [16]. HyGAS therefore adopts a spectrally matched background strategy, selecting plume-free pixels whose continuum spectra are most similar to the continuum spectra under the plume footprint. This approach aims to represent the same mixture of surface types and illumination conditions under the plume, improving representativeness of clutter estimates and stability of multi-sensor comparisons.

### *D. Scale-aware plume segmentation for multi-sensor consistency*

IME and flux estimation require a plume mask, yet plume delineation is often a dominant source of subjectivity. HyGAS implements a semi-automatic segmentation procedure designed for multi-sensor comparisons: it uses robust background statistics for thresholding and applies morphology using parameters defined in metric units and automatically rescaled to pixel units using the instrument GSD. This yields plume polygons suitable for IME integration and for consistent reporting across sensors with different spatial resolutions. The operational motivation aligns with the broader point-source quantification literature, where segmentation choices directly affect plume mass and flux [3].

Where direct matchups exist (e.g., near-simultaneous acquisitions), HyGAS supports comparison conditioning by evaluating thresholds and segmentation on the common overlap footprint, reducing domain-induced bias when comparing plume geometry and IME across missions [24], [25].

### *E. IME and emission-rate estimation in the same chain*

Once a plume mask is defined, HyGAS computes Integrated Mass Enhancement (IME) and converts IME to an emission rate using an established IME-based framework for point-source quantification from fine-scale plume observations [3]. Uncertainty is propagated from per-pixel enhancement uncertainty to IME uncertainty (by aggregating variances within the plume mask), and then to flux uncertainty by combining IME-driven and wind-driven terms in quadrature [3].

## IV. Typical HyGAS end-to-end workflow (from input to decision-ready outputs)

HyGAS is structured to support repeatable operational runs, either for single scenes (investigation) or for collections of scenes (monitoring/comparative studies). A typical workflow is:

1. **Data ingestion**: Level-1 radiance ingestion for PRISMA/EnMAP/Tanager-1 [4], [5], [8], or ingestion of Level-2 enhancement products for EMIT/GHGSat [6], [7], [20].
2. **Methane enhancement retrieval (Level-1 case)**: run MF retrieval with a chosen configuration (e.g., CTMF for heterogeneous scenes [21]; column-wise options for pushbroom artefacts [13], [18]); optionally run multiple configurations to bound sensitivity to retrieval settings and enhancement spectra [15].
3. **Uncertainty layers**: propagate instrument noise and estimate clutter; assemble total $\sigma(\Delta X)$ [17], [18]; ingest uncertainty fields for Level-2 products where provided [20].
4. **Segmentation**: apply robust thresholding and scale-aware morphology; for matchups, condition on the common overlap footprint to improve comparability [3], [24], [25].
5. **IME + flux + uncertainty**: compute IME and invert to flux using an IME-based method; combine IME and wind uncertainty (e.g., ERA5) for flux uncertainty reporting [3], [26].
6. **Outputs** include enhancement maps, per-pixel uncertainty layers, plume masks/polygons, per-plume IME and σ(IME), per-plume flux and σ(flux), and optional multi-configuration spreads for robustness assessment [3], [15].

## V. Sample outputs (PRISMA, EnMAP, Tanager-1)

### *A. Tanager-1 sample output (CTMF): Ho Chi Minh City landfill*

We report a representative HyGAS end-to-end Tanager-1 retrieval over the main landfill area in Ho Chi Minh City (Vietnam), demonstrating automated detection, segmentation, and quantification from a single acquisition. The analysed scene corresponds to the Tanager-1 product with ID 20250407_035509_25_4001 (7 April 2025 03:55:09 UTC, timestamp encoded in the product ID). HyGAS processes Level-1 radiance through MF enhancement mapping (single-image configuration), followed by scale-aware plume segmentation, IME integration, and flux inversion.

For operational robustness, we report results from the CWCMF configuration. One dominant plume is segmented and propagated through the IME/flux chain. The resulting plume extent is 18.86 km² (18,862,057 m², 17,520 pixels), corresponding to an effective ground sampling of ~33 m inferred from the georeferenced pixel area.

Flux inversion uses the wind configuration set in a HyGAS reproducible python notebook ($u_{10}$ = 2.5 m $s^{-1}$, $\sigma(u_{10})$=1.0 m $s^{-1}$, treated as the dominant atmospheric driver uncertainty). For this plume, HyGAS returns:

- IME = 6,601.64 ± 38.81 kg
- Flux = 8.89 ± 2.03 t $h^{-1}$

The uncertainty budget is dominated by the wind term ($\sigma_{Flux_wind} \approx 2.02$ t $h^{-1}$ in the run report), while the instrument-driven IME term is comparatively small for this scene

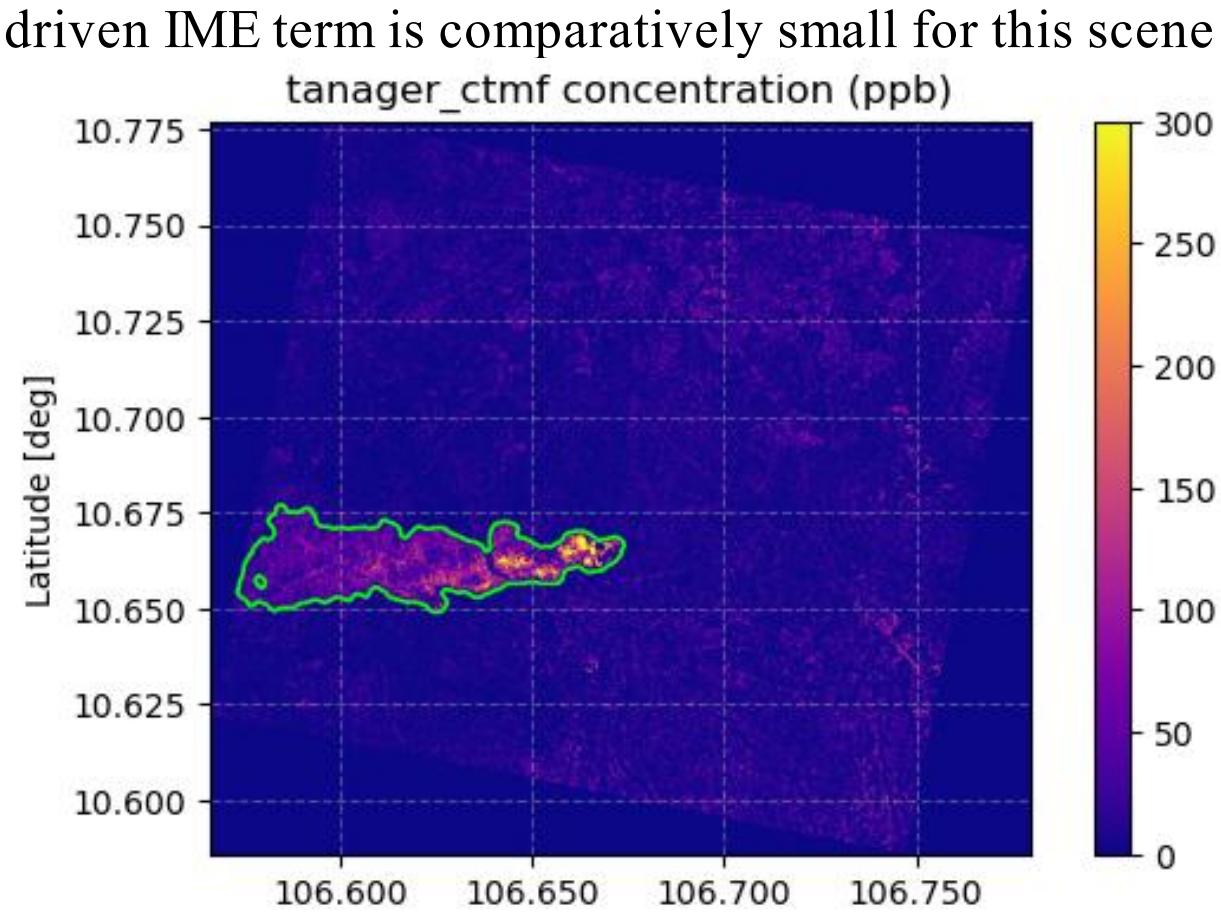

Fig. 1. Tanager-1 sample output (Ho Chi Minh City landfill, 7 April 2025). Methane enhancement map from HyGAS single-image matched filtering (CWCMF configuration). The overlaid contour indicates the final segmented plume mask used for IME integration and flux estimation; the panel summary reports plume area, IME, and flux with propagated uncertainty.

## B. EnMAP sample output

We report a representative HyGAS end-to-end EnMAP result over the Complejo Ambiental Norte III landfill (Buenos Aires, Argentina), processed from EnMAP Level-1B radiance to methane enhancement and emission rate. The analysed acquisition is 9 March 2024.

Methane enhancement $\Delta X$ is retrieved using the column-wise CWCMF configuration, in which the target and background statistics are estimated independently per detector column. This option is adopted in HyGAS as a robust operational default for pushbroom imaging spectrometers because it reduces column-dependent artefact leakage into the MF response and yields cleaner backgrounds for subsequent plume delineation.

Plumes are delineated with the HyGAS scale-aware segmentation framework (baseline physical parameters rescaled to the EnMAP native sampling), operating here at a pixel size of 31.64 m. Under the automatic thresholding and morphological constraints used in this experiment, one dominant plume is segmented with area 65.38 $km^2$ (65,378,710.51 $m^2$; approx. 65,308 pixels).

For flux inversion, 10 m wind speed is taken from ERA5-Land at acquisition time ($U_{10}$ = 2.68 m $s^{-1}$, with $\sigma(U_{10})$=1.0 m $s^{-1}$ assumed in this run) and mapped to an effective wind speed using the LES-calibrated parameterisation adopted in the HyGAS workflow.

The CWCMF returns, for the main plume:

- IME = 98,858.52 ± 481.67 kg
- Flux = 74.02 ± 16.29 t $h^{-1}$

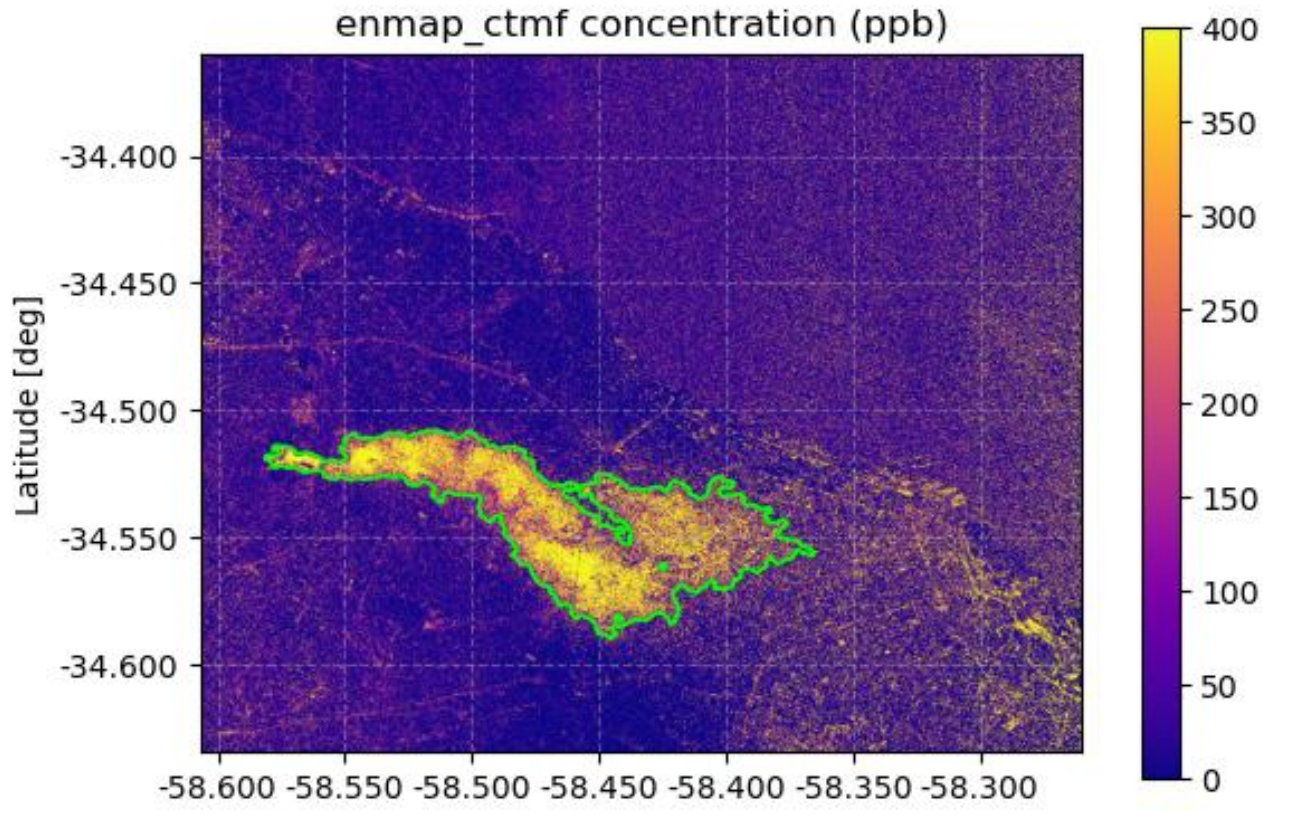

Fig. 2. EnMAP sample output (Buenos Aires, 9 March 2024). Methane enhancement field $\Delta X$ over the Complejo Ambiental Norte III landfill retrieved with HyGAS using the column-wise CWCMF configuration. The overlaid contour indicates the final segmented plume mask used for IME integration and flux estimation.

## C. PRISMA sample output

We report a representative HyGAS end-to-end PRISMA retrieval over Hassi Messaoud (Algeria), demonstrating single-scene detection, segmentation, and quantification from a MF methane enhancement product. The analysed acquisition from 24 Dec 2022, processed in HyGAS using a column-wise SRF-based matched filter (CW-CMF) in the methane window 2100–2450 nm. Plumes are delineated from the enhancement field using the same HyGAS semi-automatic, scale-aware segmentation workflow (here applied to the scaled MF output to robustly capture plume extent across the dynamic range).

Two plumes are detected in this scene. The largest plume covers 2.57 $km^2$ (2,569,892 $m^2$, 2,855 pixels), corresponding to an effective ground sampling of approximately 30 m inferred from the georeferenced pixel area. For this plume HyGAS returns:

- IME = 820.91 ± 19.82 kg
- Flux = 3.34 ± 0.69 t $h^{-1}$

Flux inversion uses the wind configuration adopted in the run ($u_{10}$ = 3.0 m $s^{-1}$, $\sigma(u_{10})$ = 1.0 m $s^{-1}$), propagated to the emission-rate uncertainty together with IME uncertainty. A secondary plume is also quantified (2.04 $km^2$, IME = 504.53 ± 15.00 kg, Flux = 2.30 ± 0.48 t $h^{-1}$), illustrating HyGAS' ability to consistently separate and report multiple emitters/plume fragments within the same PRISMA scene.

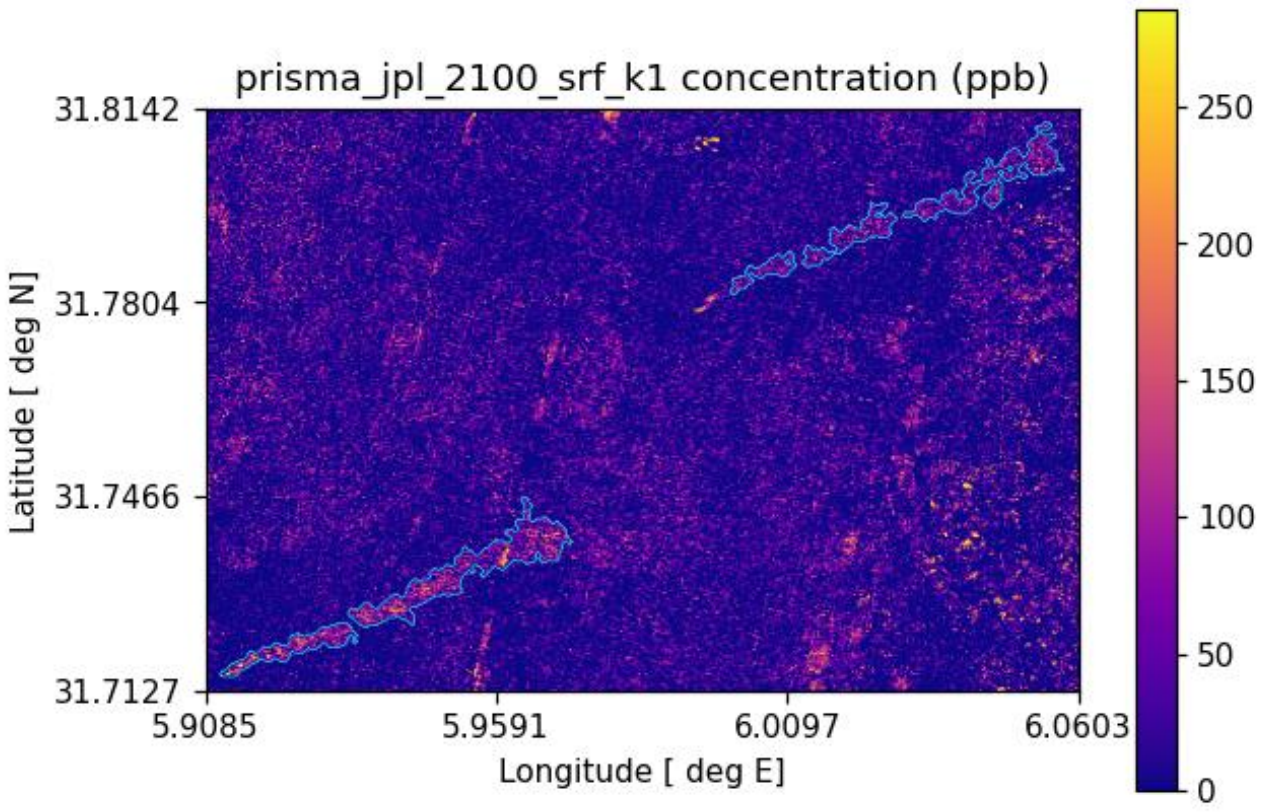

Fig. 3. PRISMA sample output (Hassi Messaoud, acquisition (24 Dec 2022). HyGAS CW-CMF methane enhancement map (2100–2450 nm; SRF-based column-wise MF). The final segmented plume polygon is overlaid on the enhancement map and used for IME integration and flux estimation; the panel summary reports plume area, IME, and flux with propagated uncertainty.

# VI. CONCLUSIONS AND OUTLOOK

HyGAS provides an open, sensor-agnostic platform for methane enhancement mapping and emission quantification that is explicitly designed for operational monitoring and comparative multi-sensor studies. It supports end-to-end Level-1 workflows for PRISMA, EnMAP, and Tanager-1 [4], [5], [8], [9], and harmonised ingestion of Level-2 methane enhancement products for EMIT and GHGSat [6], [7], [20], enabling a shared downstream chain for segmentation, IME, and flux estimation [3].

A central advantage is the explicit, end-to-end treatment of uncertainty, grounded in instrument noise models and propagated through to flux alongside wind uncertainty, supporting more standardised intercomparisons [3], [17]. Near-term extensions include strengthening sensor-specific noise characterisation (leveraging calibration/performance analyses where available [14], and mission-specific noise coefficient retrieval approaches [18]) and expanding the set of validated multi-sensor matchups using published methane mapping baselines over EnMAP and PRISMA [24], [25].

# ACKNOWLEDGMENT

This research was supported by the Italian Space Agency ASI within the CLEAR-UP project (Contract/Agreement n. 2022-16-U.0; CUP n. F83C22000780005). We acknowledge Planet for providing access to Tanager-1 Core Imagery data used in this study. We also gratefully acknowledge the German Aerospace Center (DLR) for providing EnMAP data.